\RequirePackage{ifpdf}
\ifpdf 
\documentclass[pdftex]{sigma}
\else
\documentclass{sigma}
\fi

\begin{document}
\allowdisplaybreaks

\renewcommand{\PaperNumber}{047}

\FirstPageHeading

\ShortArticleName{Internal Modes of Solitons and Near-Integrable
Highly-Dispersive Nonlinear Systems}

\ArticleName{Internal Modes of Solitons and Near-Integrable\\
Highly-Dispersive Nonlinear Systems}

\Author{Oksana V. CHARKINA and Mikhail M. BOGDAN}

\AuthorNameForHeading{O.V. Charkina and M.M. Bogdan}

\Address{B. Verkin Institute for Low Temperature Physics and Engineering of the NAS of Ukraine,\\
47 Lenin Ave., Kharkiv, 61103 Ukraine}

\Email{\href{mailto:charkina@ilt.kharkov.ua}{charkina@ilt.kharkov.ua},
 \href{mailto:bogdan@ilt.kharkov.ua}{bogdan@ilt.kharkov.ua}}

\ArticleDates{Received November 30, 2005, in f\/inal form April
11, 2006; Published online April 28, 2006}

\Abstract{The transition from integrable to non-integrable
highly-dispersive nonlinear models is investigated. The
sine-Gordon and $\varphi^4$-equations with the additional
fourth-order spatial and spatio-temporal derivatives, describing
the higher dispersion, and with the terms originated from
nonlinear interactions are studied. The exact static and moving
topolo\-gical kinks and soliton-complex solutions are obtained for
a special choice of the equation parameters in the dispersive
systems. The problem of spectra of linear excitations of the
static kinks is solved completely for the case of the regularized
equations with the spatio-temporal derivatives. The frequencies of
the internal modes of the kink oscillations are found explicitly
for the regularized sine-Gordon and $\varphi^4$-equations. The
appearance of the f\/irst internal soliton mode is believed to be
a criterion of the transition between integrable and
non-integrable equations and it is considered as the
suf\/f\/icient condition for the non-trivial (inelastic)
interactions of solitons in the systems.}


\Classification{34A05; 34A34; 35G25}

\bigskip

\rightline{\it We dedicate this paper to the memory of
A.M.~Kosevich.}

\section{Introduction}

The soliton theory of completely integrable systems  proposed a
new basis of nonlinear fundamental excitations \cite{AbSeg}. These
elementary waves are solitons (antisolitons), breathers
(soliton-antisoliton bound states) and linear waves of the
continuous spectrum. Any localized initial condition evolves as a
superposition of such basic excitations. In the case of  {\it
trivial} (elastic) interaction between solitons the nonlinear
excitations undergo only the phase and mass center shifts as a
result of pair collisions. Then  the initial prof\/ile is
transformed asymptotically in the long time limit, in general,
into a sequence of solitons and breathers existing upon  linear
wave background.

Soliton interactions can be described explicitly by the use of a
multisoliton formula. To study interaction between a soliton and
linear waves one needs to solve the equation linearized near the
given soliton solution. It is remarkable that an exact solution of
the problem can be also obtained  from the multisoliton formula.
For this purpose it is enough to make a special choice of
parameters in the formula, which provides specif\/ication of the
soliton and a near-zero amplitude limit for breather excitations
\cite{DasHN}.

In this paper we give a short survey of stability properties and
linear excitation spectra of solitons in the sine-Gordon and
$\varphi^4$-equations and their Boussinesq-like generalizations.
Then we present a statement about  spectrum of linear excitations
of a soliton in a completely integrable system with  trivial
interaction between solitons. We justify the hypothesis which
contends that  presence of the internal oscillation mode in a
soliton spectrum serves as a~suf\/f\/icient condition of
non-integrability of a nonlinear system \cite{BKV}. We use this
criterion to state a crossover from the integrable sine-Gordon
equation to its non-integrable generalizations with higher-order
derivatives and dispersive nonlinear terms as well to conclude
about non-integrability of corresponding $\varphi^4$-systems.
These new equations possess exact moving kink or soliton complex
solutions but they appear to be non-integrable.

\section[Kink excitation spectra in the sine-Gordon and $\varphi^4$-systems]{Kink
excitation spectra in the sine-Gordon and
$\boldsymbol{\varphi^4}$-systems}

One of most important problems concerning the soliton dynamics in
integrable and non-in\-teg\-rab\-le systems is stability
investigation. To solve the problem analytically one has to f\/ind
a~spectrum of linear excitations of the soliton. Usually this
approach is demonstrated by consideration of the sine-Gordon
equation (SGE) as an example:
\begin{gather*}
u_{tt}-u_{xx}+\sin u=0.
\end{gather*}
The well-known moving kink solution of the SGE has the following
form:
\begin{gather*}
u_s=4\arctan\left(\exp{x-Vt\over\sqrt{1-V^2}}\right).
\end{gather*}
Due to the Lorentz invariance of the SGE it is enough to determine
a spectrum of kink excitations of a static solution $u_s$ with
$V=0$. For the amplitude of small deviations from a kink prof\/ile
\begin{gather*}
u-u_s=\psi(x)\exp{i\omega t}
\end{gather*}
it is easy to obtain the linear Schr\"{o}dinger-type equation with
the simplest ref\/lectionless potential~\cite{LL}:
\begin{gather*}
\left(-{d^2\over dx^2}+1-{2\over\cosh ^2 x}\right)\psi(x)=\omega
^2\psi(x).
\end{gather*}
Eigenfunctions of the equation correspond to the translational
mode with the zero eigenfrequency
\begin{gather*}
\psi_0(x)=1/\cosh x, \qquad \omega_0=0
\end{gather*}
and to waves of the continuous spectrum:
\begin{gather*}
\psi_k(x)=(\tanh{x}+ik)\exp{ikx}, \qquad \omega_k=\sqrt{1+k^2}.
\end{gather*}
No instability mode with $\omega^2<0$ exists in the spectrum hence
the kink in the integrable SG-system is stable. We note also that
there is no additional localized mode with a discrete eigenvalue
lying in the frequency gap $0<\omega<1$.

The another situation takes place in the non-integrable
$\varphi^4$-equation:
\begin{gather*}
\varphi_{tt}-\varphi_{xx}-2\big(\varphi-\varphi^3\big)=0.
\end{gather*}
In this case there is also an exact kink solution:
\begin{gather*}
\varphi_s=\tanh{x-Vt\over\sqrt{1-V^2}}.
\end{gather*}
Then for the amplitude $f(x)$ of small deviations from a static
kink
\begin{gather*}
\varphi-\varphi_s=f(x)\exp{i\omega t}
\end{gather*}
one f\/inds the linear equation with one more ref\/lectionless
potential:
\begin{gather*}
\left(-{d^2\over dx^2}+1-{6\over\cosh ^2 x}\right)f(x)=\omega
^2f(x).
\end{gather*}
However, besides the translational mode with the zero frequency:
\begin{gather*}
f_0(x)=1/\cosh^2 x, \qquad \omega_0=0
\end{gather*}
and continuum waves $f_k(x)$ with $\omega_k=\sqrt{4+k^2}$, the
eigenfrequency spectrum of the equation contains the additional
discrete eigenvalue corresponding to the {\it internal mode} of
kink oscillations:
\begin{gather*}
f_1(x)={\sinh x\over\cosh^2x}, \qquad \omega_1=\sqrt{3}.
\end{gather*}
Since all $\omega^{2}_i\geq 0$ then the kink of the
$\varphi^4$-equation is also stable. But in contrast to the SG
kink it possesses an intrinsic structure giving rise to an
internal degree of freedom. The internal mode corresponds to a
localized oscillation of an ef\/fective width of the kink.

Analogous situation takes place in the double sine-Gordon
equation~\cite{cam2}
\begin{gather*}
u_{tt}-u_{xx}+ \sin u \cos u+h\sin u=0,
\end{gather*}
where the internal mode of the wobbler solution describes
antiphase oscillations of two composite kinks. In this case the
frequency of the internal mode becomes a function of the
parameter~$h$.
 Numerical simulations of  the double sine-Gordon and
$\varphi^4$-equations demonstrate ef\/fects of inelastic
interactions between solitons, including resonant phenomena
\cite{cam2,cam1}. It is natural to suggest that the appearance of
the internal mode in these models is connected with a
non-integrability of the systems. This question is analyzed in the
next section.

\section{Existence of internal mode as a non-integrability criterion}

The above consideration of soliton excitations allows to formulate
the following statement: {\it in a~completely integrable system
with only trivial (elastic) interactions between solitons a
spectrum of linear excitations for a soliton can consist only of a
translational (Goldstone) mode and continuum waves.} Here and
further we use the terminology of {\it completely integrable
systems}, {\it trivial} and {\it non-trivial} interactions
following those of V.E.~Zakharov's papers (see,
e.g.,~\cite{BZakh}). Well-known examples of the completely
integrable systems with trivial soliton interactions are the
sine-Gordon, nonlinear Schr\"{o}dinger and Landau--Lifshitz
equations. In such systems a basis of fundamental excitations
consists of solitons, breathers and linear waves. To solve a
spectral problem for the equations linearized near a soliton
solution one can use a multisoliton formula. Then a soliton
solution with arbitrary small perturbations can be presented by a
combination of a soliton and breathers with near-zero amplitudes.
In a small amplitude limit a breather is delocalized and
transformed into a linear wave with a frequency belonging to a
continuous spectrum. No {\it linear} localized mode corresponding
to a discrete eigenvalue in the frequency gap can be obtained from
this basis of fundamental excitations. Exception is the existence
of a~translational mode which follows from the translational
invariance of the systems. Hence the spectrum of linear
excitations of the soliton in the case of the trivial soliton
interaction consists of the translational (Goldstone) mode and
continuum waves. Thus it appears that solitons in a completely
integrable system with only trivial interactions between solitons
have no internal modes in their spectra of linear excitations. And
vice versa, presence of the internal mode in a~soliton spectrum
should be considered as a suf\/f\/icient condition at least for
existence of non-trivial interaction between solitons in 
a~nonlinear system. The last statement can be strengthened and
formulated as the following conjecture.

\medskip

\noindent
 {\bf Hypothesis.}
 {\it Presence of the internal mode
in a spectrum of linear excitations of a soliton is a~sufficient
condition of the non-integrability of corresponding
soliton-bearing nonlinear equations.}

\medskip

It should be noted that the absence of the internal mode in a
linear excitation spectrum of a soliton is not a suf\/f\/icient
condition of the integrability of the soliton-bearing system. The
main argument in favour of the hypothesis is that in the
completely integrable system only the breather has its frequency
in the gap between the zero and the continuous spectrum due to the
f\/inite value of its amplitude. Hence the existence of another
type of  nonlinear localized excitation with  frequency in the gap
would be assumed to obtain the internal mode oscillation as its
linear limit. Such kind of soliton excitations is not known yet,
at least for integrable systems.

The hypothesis f\/irst was formulated in the paper~\cite{BKV} and
used for investigation of the transition from the integrable
Landau--Lifshitz equation to a non-integrable one describing the
biaxial ferromagnet placed in a magnetic f\/ield. Later this idea
was used in the calculation of  detachment of the internal mode
from the continuous spectrum in the double sine-Gordon and
near-discrete nonlinear Schr\"{o}dinger equations~\cite{Kivsh}.

\section{Soliton motion and instabilities in highly dispersive systems}

A soliton motion in the above examples of integrable equations
does not inf\/luence the soliton stability. It is not a common
case for integrable systems. There are known integrable systems
with {\it nontrivial} interactions between solitons which exhibit
instabilities in soliton dynamics \cite{Berrym,FST,KSF}. The most
famous example is the Boussinesq equation
\cite{BZakh,Yajima,FPR,TajMur}:
\begin{gather*}
u_{tt}-u_{xx}-6\big(u^2\big)_{xx}-u_{xxxx}=0.
\end{gather*}
It is easy to see that in this equation even the trivial solution
$u=0$ is unstable with respect to short waves $u=a\exp(\gamma
t)\cos kx$ of a continuous spectrum with wave numbers $k>1$. An
exact analysis~\cite{Berrym} of linear stability of a soliton of
the Boussinesq equation also indicates the existence of an
instability of the solution. Nevertheless it was shown that a
nonlinear stage of the instability can be described by almost
independent evolution of the growing mode \cite{Yajima} which
destroys really neither solitons nor breathers~\cite{TajMur}.

However, there is another channel of the soliton instability
directly connected with the nontrivial interaction between
solitons in the Boussinesq equation with the opposite sign of the
fourth-order derivative term~\cite{FPR}. Recently the Boussinesq
equation was revisited and investigated in detail by the dressing
method~\cite{BZakh}. As a result of the consideration, a soliton
decay into a pair of composite solitons (or its collapse) had been
described explicitly. Linear stability analysis of the decaying
soliton at the initial stage of this inelastic process has to lead
to an existence of the {\it instability} mode \cite{FST} which
corresponds to a localized eigenfunction. Here we should emphasize
that we distinguish this instability mode corresponding to a
growing solution from the internal mode which oscillates in time.
To our knowledge, the internal mode has not been found yet in
integrable equations with the non-trivial soliton interaction.

Taking into account higher-order derivatives to describe strong
dispersion in the sine-Gordon and double sine-Gordon equations one
comes to the highly dispersive non-integrable models~\cite{BKM}.
These equations possess the soliton-complex solutions which can
propagate with the f\/ixed velocity. For example the dispersive
sine-Gordon equation
\begin{gather*}
u_{tt}-u_{xx}-\beta u_{xxxx}+\sin u=0
\end{gather*}
has the $4\pi$-soliton complex solution in the form
\begin{gather}
u_c=8\arctan[\exp\varepsilon_{0}(x-V_{0}t)]\label{eq16},
\end{gather}
where $\varepsilon_{0}$ and $V_0$ are def\/inite functions of the
dispersive parameter $\beta$. The soliton complex is a~specif\/ic
bound state of identical solitons, which is formed due to  strong
dispersion (see~\cite{BKM} for a survey). Note that a highly
dispersive system is not integrable if this kind of the solution
is revealed in the corresponding dispersive equations.

To avoid an instability with respect to short waves the
regularized dispersive sine-Gordon equation with a mixed
fourth-order derivative was introduced \cite{BKM}:
\begin{gather}
u_{tt}-u_{xx}-\beta u_{ttxx}+\sin u=0\label{eq17}.
\end{gather}
It also has the exact soliton-complex solution of the same form
\eqref{eq16} but with  dif\/ferent velocity dependence on the
dispersive parameter $\beta$, namely:
\begin{gather*}
V_{r}=\sqrt{1+\frac \beta 3}-\sqrt{\frac \beta 3}.
\end{gather*}

Dynamical properties of the $2\pi$-kink and the $4\pi$-soliton
complex in the regularized  equation~\eqref{eq17} have been
studied in detail. In particular a spectrum of linear excitations
of the static kink in the regularized sine-Gordon equation has
been found exactly~\cite{CB}. It was shown that, depending on the
$\beta$ value, one or more internal modes are present in the
spectrum. Additionally for the regularized dispersive
$\varphi^4$-equation a complete spectrum of internal modes of kink
oscillations was found exactly. Thus one of the advantages of the
regularized dispersive equations is the fact that a process of the
appearance of the internal modes of static kinks can be described
explicitly.

It should be noted that there are several mechanisms of formation
of multikinks and soliton bound states. At f\/irst it was found
that interactions of oscillating soliton tails in dispersive media
lead to the formation of bunches of solitons, consisting of two or
more well-distinguishable humps~\cite{GorOs,Kaw}. Later it has
been shown that solitons coexisting with resonant radiation can
form bound states with purely solitonic asymptotics due to some
kind of the radiation interference ef\/fect \cite{Cham1,Cham2}.
These multisoliton bound states have been called embedded
solitons, implying their existence in the continuous spectrum.

In the paper [15] the concept of the soliton-complex in a
nonlinear dispersive medium was proposed. It was shown that
solitons in a strongly dispersive medium possess an internal
structure and their interaction depends on intrinsic properties
such as f\/lexibility. Due to this dependence, the potential
energy turns out to be a non-monotonic function of the distance
between solitons. As a result, even identical topological solitons
can attract each other and form a bound soliton complex which can
move without any radiation in strongly dispersive media. These
bound soliton states with zero and small distances between
composite solitons have been called the soliton complexes. Their
formation cannot be described by the soliton perturbation theory.
Such soliton complexes appear even in systems with the ef\/fective
strong dispersion, for example in nonlinear Schr\"{o}dinger
equation with the parametric pumping and dissipation \cite{Bar}.
The soliton complex in this case is formed due to neither
oscillatory soliton tails interaction nor embedded soliton
interaction but as a result of a strong multi-soliton interaction.

The regularized equations give a possibility to study soliton
complexes in conditions when the inf\/luence of radiation is
minimized. Indeed, in the equation \eqref{eq17} the continuous
spectrum of linear waves degenerates into a single frequency
$\omega_{0}=1$ for the parameter $\beta=1$. At the same time the
number of kink internal modes become inf\/inite and they play a
principal role in soliton-complex dynamics.

More than thirty years ago A.~Kosevich and A.~Kovalev combined the
sine-Gordon and the modif\/ied Boussinesq equations to describe
the crowdion (dislocation) motion in crystal with nonlinear
interaction between nearest atoms \cite{KK}. They showed that the
equation (KKE)
\begin{gather*}
u_{tt}-u_{xx}+\sin u-\gamma u^2_xu_{xx}-\beta
u_{xxxx}=0
\end{gather*}
has a moving crowdion solution with an arbitrary value of the
velocity parameter $V$
\begin{gather*}
u_{cr}=4\arctan[\exp\varepsilon(x-Vt)],
\end{gather*}
if the relation $\gamma={3\beta /2}$ takes place. Here the
dependence of the parameter $\varepsilon$ on the velocity is the
following:
\begin{gather*}
\varepsilon=\sqrt{{1-V^2\over 2\beta}\left(\sqrt{1+{4\beta\over
(1-V^2)^2}} -1\right)}.
\end{gather*}
Besides, for the equation, combining the Boussinesq and
$\varphi^4$-equations,
\begin{gather}
\varphi_{tt}-\varphi_{xx}-2(\varphi-\varphi^3)-\beta\varphi_{xxxx}+\alpha
\varphi_x\varphi_{xx}=0\label{eq21}
\end{gather}
authors showed the existence of a moving kink solution with a
def\/inite velocity which is a~function of a special combination
of parameters $\beta$ and $\alpha$.

A year later after the paper \cite{KK} K.~Konno {\it et al}
derived and solved by the inverse scattering method the
equation~\cite{Kon}:
\begin{gather}
u_{xt}-\sin u+\beta\left(u_{xxxx}+{3\over 2}
u^2_xu_{xx}\right)=0.\label{eq22}
\end{gather}
This equation was reduced by the use of the Hirota transformation
\begin{gather}
u=2i\ln{f^*\over f}\label{eq23}
\end{gather}
to the following bilinear equations (see, e.g., \cite{ChDZ})
\begin{gather}
\big(D_xD_t+\beta D_x^4\big)f\cdot f={1\over
2}\big(f^2-f^{*2}\big)\label{eq24},
\\
D_x^2 f\cdot f^*=0\label{eq25},
\end{gather}
where Hirota operators are def\/ined as follows \cite{Hiro}
\begin{gather*}
D^m_x D^n_t a\cdot b=(\partial_x-\partial_{x'})^m
(\partial_t-\partial_{t'})^n a(x,t)b(x',t') \big|
_{x=x',t=t'}.
\end{gather*}
The bilinear equations \eqref{eq24} and \eqref{eq25} are solved
exactly, resulting in the multisoliton solution.

In conclusion of the section we would like to note that only
additional integrals of motion cannot guarantee, obviously,
complete integrability of nonlinear equations. Here we mention,
for example, the near-integrable dispersive equation by Leo {\it
et al} \cite{LLS}:
\begin{gather*}
u_t-u_x\sin u+\beta u_{xxx}=0,
\end{gather*}
which has three independent integrals of motion but the
quasi-elastic soliton interaction in this equation is conf\/irmed
only numerically yet.

\section{Kinks and soliton complexes in highly dispersive models\\
with anharmonic interatomic interactions}

Instead of the use of a variational approach \cite{BZ} for
studying systems with anharmonic interatomic interactions, in this
section we seek for exact solutions of the regularized versions of
the dispersive Boussinesq-like sine-Gordon and
$\varphi^4$-equations. However f\/irst of all we would like to
conclude about the integrability of the Kosevich--Kovalev
equation. For the case $\gamma=3\beta/2$ it has remained an open
problem. There was a premature conclusion about its
integrability~\cite{GM}, apparently based on  similarity of this
special case of the KKE and equation \eqref{eq22}. We use the
Hirota transformation \eqref{eq23} to reduce the KKE to a set of
bilinear equations
\begin{gather*}
\big(D_t^2-D_x^2-\beta D_x^4\big)f\cdot f={1\over
2}\big(f^2-f^{*2}\big),
\\
D_x^2 f\cdot f^*=0.
\end{gather*}
It is easy to verify that these equations do not have a
multisoliton solution in contrast to equations \eqref{eq24} and
\eqref{eq25}. Nevertheless in the case $\gamma\neq3\beta/2$ we
f\/ind the exact soliton-complex solution of the KKE:
\begin{gather*}
u_c=8\arctan[\exp\varepsilon(x-Vt)], \qquad \varepsilon^4={1\over
3\beta-8\gamma},\qquad V=\sqrt{1-2{\beta-4\gamma\over
\sqrt{3\beta-8\gamma}}}.
\end{gather*}
Evidently the solution exists when $\beta > 8\gamma /3$ and the
velocity $V<1$ if $\beta>4\gamma$.

Now we consider the {\it regularized} dispersive sine-Gordon
equation with a nonlinear dispersive term:
\begin{gather}
u_{tt}-u_{xx}+\sin u-\gamma u^2_xu_{xx}-\beta
u_{ttxx}=0\label{eq33}.
\end{gather}
At f\/irst we show that this equation has a static kink solution
for arbitrary values of parameters~$\gamma$ and $\beta$. In fact
after one integration of the equation \eqref{eq33} we obtain for
$u(x)$
\begin{gather*}
du/dx=\left[\left(\sqrt{1+8\gamma\sin^{2}(u/2)}-1\right)/\gamma\right]^{1/2}.
\end{gather*}
The $2\pi$-kink solution exists for any positive $\gamma$ and at
small values of the parameter we f\/ind out the expansion
\begin{gather}
u_{k}(x)\thickapprox 4\arctan(\exp(x))-2\gamma {\sinh(x)\over
\cosh^{2}(x)}.\label{eq35}
\end{gather}
With increasing $\gamma$ the ef\/fective kink width also
increases. We have analyzed a spectral problem of the equation
linearized near the kink solution  \eqref{eq35}. By the use of the
perturbation theory with a small parameter $\gamma$ we have
revealed  detachment of the internal mode from the continuous
spectrum. We have found that a dif\/ference of the internal mode
frequency from unity is proportional to the f\/irst power of the
parameter $\gamma$. Thus we conclude that the
equation~\eqref{eq33} is not integrable.

Besides the static kink the equation \eqref{eq33} possesses the
soliton-complex solution
\begin{gather*}
u_c=8\arctan[\exp\kappa(x-Vt)],
\end{gather*}
where parameters $\kappa$ and $V$ obey the following equations
\begin{gather*}
\kappa^4={1\over 3\beta V^{2}-8\gamma}, \qquad
V^2={{\kappa^2-1}\over \kappa^2 (1-\beta \kappa^{2})}.
\end{gather*}
The latter equations determine f\/ixed velocity and ef\/fective
width of the soliton complex. Although there is the lower boundary
for admissible values of the velocity ($V>\sqrt{8\gamma/3\beta}$)
nevertheless at small enough $\gamma$ there is a def\/inite range
of $\beta$-values which provides valid quantities of parameters
$\kappa$ and $V$. For example, for $\beta=0.03$ and $\gamma=0.025$
values of the velocity $V_0$ and the parameter $\kappa_0$  are
equal to $0.913$ and $1.47$ respectively.

We propose one more regularized version of the dispersive
Boussinesq-like sine-Gordon equation
\begin{gather}
u_{tt}-u_{xx}+\sin u-\gamma u^2_xu_{tt}-\beta
u_{ttxx}=0\label{eq38}.
\end{gather}
The equation possesses the moving kink and soliton complex
solutions which are very similar in their forms to corresponding
solutions of the KKE. In the case $\gamma={3\beta /2}$ there
exists the $2\pi$-kink propagating with an arbitrary velocity less
than unity:
\begin{gather*}
u_s=4\arctan[\exp\varepsilon(x-Vt)], \qquad
\varepsilon=\sqrt{{1-V^2\over 2\beta V^2}\left(\sqrt{1+{4\beta
V^2\over (1-V^2)^2}}-1\right)}.
\end{gather*}
Such  moving solution does not exist in the case of
$\gamma\neq{3\beta /2}$. However a static $2\pi$-kink exists for
arbitrary values of $\beta$ and $\gamma$:
\begin{gather}
u_k=4\arctan(\exp(x)).\label{eq41}
\end{gather}
The kink exhibits an intrinsic structure which manifests itself in
its internal dynamics and complex soliton interactions as it is
shown below.

For values $\gamma\neq{3\beta /2}$ the $4\pi$-soliton complex
solution appears:
\begin{gather}
u_c=8\arctan[\exp\varepsilon(x-Vt)],\label{eq42}
\\
V={\sigma\over \sqrt{1+\sigma^2}+1},\qquad
\varepsilon=\sqrt{{\sqrt{1+\sigma^2
}+1\over\sigma\sqrt{3\beta-8\gamma}}},\nonumber
\end{gather}
where
\begin{gather*}
\sigma={\sqrt{3\beta-8\gamma}\over\beta-4\gamma}, \qquad
\beta>4\gamma.
\end{gather*}
It should be noted that this solution exists also for $\beta=1$
when the continuous spectrum dege\-ne\-rates and it is interesting
to investigate numerically the soliton-complex stability
properties in this case.

\begin{figure}[t]
\centerline{\includegraphics[height=6cm]{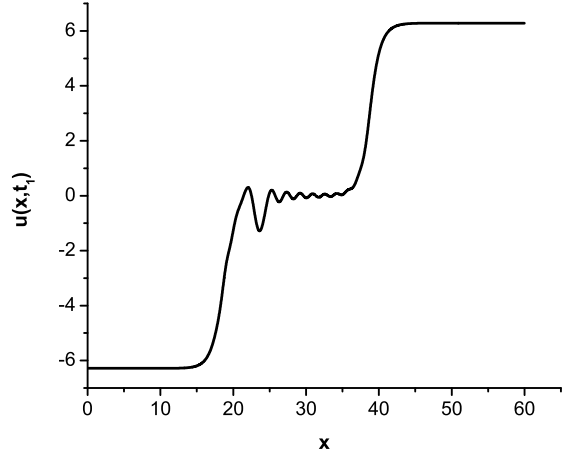}}
\vspace{-2mm} \caption{Dissociation of the soliton complex
prof\/ile with the initial velocity $V_{\rm in}=0.4$ in the
dispersive sine-Gordon equation \eqref{eq38} with $\beta=1$ and
$\gamma=0.025$. The numerical solution prof\/ile is shown at the
moment $t_{1}=60$.} \label{fig1}
\end{figure}

Here we present results of numerical simulations of the equation
\eqref{eq38} for the soliton-complex dynamics when the parameter
$\beta=1$ and $\gamma$ is very small. If one starts at $t_{0}=0$
with the soliton-complex prof\/ile which has the initial velocity
$V_{\rm in}=0.4$ less than $V_{r}$ then the result of simulation
will be as shown in Fig.~\ref{fig1}. The prof\/ile evolves with
throwing down the superf\/luous energy by exciting the internal
mode, and it dissociates f\/inally into two moving kinks.

The soliton complex given by the exact solution \eqref{eq42}
appears to be stable and moves radiationlessly in the dispersive
media.

In conclusion of this section we present the regularized version
of the generalized dispersive $\varphi^4$-equation \eqref{eq21} of
the following form:
\begin{gather*}
\varphi_{tt}-\varphi_{xx}-2\big(\varphi-\varphi^3\big)-\beta\varphi_{xxtt}+\alpha
\varphi_x\varphi_{tt}=0.
\end{gather*}
Besides the moving kink which is analogous to the solution of the
equation \eqref{eq21} there exists an exact static kink solution
for arbitrary parameters $\beta$ and $\alpha$:
\begin{gather*}
\varphi_s(x)=\tanh(x).
\end{gather*}
In the next section explicit expressions for frequencies of
internal modes of static kinks are obtained for the regularized
dispersive sine-Gordon and $\varphi^4$-equations.

\section[Internal modes of kinks in the dispersive sine-Gordon and
$\varphi^4$ equations]{Internal modes of kinks in the dispersive
sine-Gordon\\ and $\boldsymbol{\varphi^4}$ equations}

A remarkable property of the regularized equation \eqref{eq38} is
that the problem of the spectrum of internal modes of a static
kink can be solved exactly.

Let us seek for a solution of equation \eqref{eq38} in the form:
\begin{gather*}
u=u_k+\psi(x)\exp{i\omega t},
\end{gather*}
where $u_k$ is given by equation \eqref{eq41} and $\psi(x)\ll
u_k$. Then the linearized equation for function~$\psi(x)$ is
obtained as follows:
\begin{gather}
\left[ -(1-\beta\omega^2){d^2\over
dx^2}+1-{2\over\cosh^2x}(1-2\gamma
\omega^2)\right]\psi=\omega^2\psi\label{eq48}.
\end{gather}
Discrete set of eigenvalues is given by the equation (see, e.g.,
\cite{LL}):
\begin{gather}
\sqrt{{1\over 4}+{2(1-2\gamma\omega^2)\over
1-\beta\omega^2}}-\sqrt{1-\omega^2\over1-\beta\omega^2}=n+ {1\over
2},\label{eq49}
\end{gather}
where $n$ is an integer. The zero eigenvalue corresponds to the
translational mode and $n=0$. The number of further levels, i.e.\
the number of internal modes, is determined by values of the
parameters $\beta$ and $\gamma$.

Let us consider some special cases of the equation \eqref{eq49}:

1. $\gamma=0$. For this case the dispersive SGE was studied in the
paper \cite{CB}:
\begin{gather*}
\sqrt{{1\over 4}+{2\over 1-\beta\omega^2}}-
\sqrt{1-\omega^2\over1-\beta\omega^2}=n+ {1\over 2}.
\end{gather*}
Depending on the dispersive parameter $\beta$, internal modes
detach from the continuous spectrum at the following values of the
dispersive parameter $\beta$:
\begin{gather*}
\beta_n=1-{2\over n(n+1)}.
\end{gather*}
The f\/irst mode frequency is presented explicitly as follows:
\begin{gather*}
\omega_1=\sqrt{{1\over\beta}{\Delta^2-9\over\Delta^2-1}}, \qquad
\Delta(\beta)={6\beta+\sqrt{17\beta^2-10\beta+9}\over 1+\beta}.
\end{gather*}
Analysis of the expansion of $\omega_1(\beta)$ at small $\beta$
indicates clearly the appearance of the f\/irst internal mode as
soon as $\beta\neq 0$ and hence  loss of the integrability of this
variant of the sine-Gordon equation.

It should be noted that in the case of the continuous spectrum
degeneration ($\beta=1$) all the inf\/inite set of the frequencies
of internal modes is expressed very simply:
\begin{gather*}
\omega_n=\sqrt{1-{2\over (n+1)(n+2)}}.
\end{gather*}
In this limit the dynamics of soliton complexes depends strongly
on possibility of excitation of the internal modes of composite
kinks. This circumstance leads to a large variety of resonant
phenomena in soliton interactions in the dispersive
equation~\eqref{eq38}.

2. $\gamma=\beta/2$. It is easy to be convinced that this is the
case of a {\it reflectionless} potential in the
equation~\eqref{eq48}
\begin{gather*}
\sqrt{1-\omega^2\over1-\beta\omega^2}=1-n.
\end{gather*}
It is evident that there are only two discrete eigenvalues: (i)
values $n=0$ and $\omega=0$ correspond to the translational mode;
(ii) values $n=1$ and $\omega=1$ correspond to the edge of the
continuous spectrum. Such kind of the spectrum is typical for
integrable systems. Unfortunately the absence of internal mode
cannot serve as the suf\/f\/icient condition of the integrability.
Therefore we doubt  complete integrability of the equation in this
case but could expect the near-integrable behavior of soliton
dynamics.

3. $\gamma=3\beta/2$. This is the case when moving kink exists.
For the static kink limit the discrete eigenvalue equation is the
following
\begin{gather*}
\sqrt{{1\over 4}+{2(1-3\beta\omega^2)\over
1-\beta\omega^2}}-\sqrt{1-\omega^2\over1-\beta\omega^2}=n+ {1\over
2}.
\end{gather*}
Here also no internal mode exists between the zero frequency and a
continuous spectrum. Howe\-ver the potential well in the equation
\eqref{eq48} appears to be not ref\/lectionless as usually in
integrable systems. The last argument works rather against the
suggestion about the equation integrability in this case.

Spectrum of internal modes of a static $\varphi^4$-kink is given
by the equation:
\begin{gather*}
\sqrt{{1\over 4}+{6+\alpha\omega^2\over
1-\beta\omega^2}}-\sqrt{4-\omega^2\over1-\beta\omega^2}=n+ {1\over
2}.
\end{gather*}
This equation has discrete set of solutions for eigenfrequencies
of internal modes for any positive~$\alpha$ and $0\leq\beta<1$. A
structure of the discrete spectrum of internal modes is similar to
that of a kink of the dispersive $\varphi^4$-equation \cite{CB}.
In the latter paper the detailed study of the kink excitations
spectrum for the case of $\alpha=0$ was performed.

Thus the analysis of  presence of internal modes in the kink
excitation spectrum justif\/ies the fact of the non-integrability
of the corresponding dispersive sine-Gordon and
$\varphi^4$-equations. Evidently, the internal structure of
solitons in highly dispersive media can manifest itself in
inelastic processes of the soliton interaction. Formation of the
soliton complexes is one of consequences of such an interaction.
Hence the existence of the soliton complex solution can be also
considered as the argument in favour of the non-integrability of
nonlinear equations.

\section{Conclusions}

The results of the present consideration are summarized as
follows:

1. Exact static and moving solutions describing kinks and soliton
complexes are found in the equations describing highly dispersive
models with nonlinear interatomic interactions. However the
presence of the exact solutions (especially, moving kinks with
arbitrary velocities) does not argue for  complete integrability
of the systems. In particular, it is demonstrated that the
Kosevich--Kovalev equation is not completely integrable in the
case $3\gamma>8\beta$ due to an existence of the soliton complex
solution. It is also shown that there is not the standard
multisoliton formula for this equation in the special case $\gamma
= 3\beta/2$.

2. As a result of application of the criterion of the
non-integrability, based on presence of the internal mode in a
soliton spectrum, it is shown that the regularized dispersive
sine-Gordon equation with the fourth-order mixed derivative and
the nonlinear dispersive term is not completely integrable for the
$\gamma<\beta/2$. Soliton dynamics in this system demonstrates
a~complex behavior including bound state formation and
dissociation processes. However it should be expected for the case
$\gamma=\beta/2$ that soliton interactions in the system could be
close with the dynamical behavior in the integrable model.

3. In general, f\/inding of the exact solutions for spectra of
linear excitations of solitons, including the internal modes, for
the generalized sine-Gordon and $\varphi^4$-models allow us to
make conclusions about the near-integrable soliton properties or a
complex internal structure and the non-trivial dynamics  in the
highly dispersive systems. Finally it is believed that existence
of the internal mode  can serve as the criterion of
non-integrability of a nonlinear system, supplementing
ef\/f\/iciently other integrability checks, like, e.g., the
Painlev\'e test.

\LastPageEnding

\end{document}